\documentclass[lettersize,journal]{IEEEtran}
\usepackage{amsmath,amsfonts,amssymb}
\usepackage{algorithmic}
\usepackage{algorithm}
\usepackage{array}
\usepackage{booktabs}
\usepackage[caption=false,font=footnotesize,labelfont=sf,textfont=sf]{subfig}
\usepackage{textcomp}
\usepackage{stfloats}
\usepackage{url}
\usepackage{graphicx}
\usepackage{cite}
\usepackage{xcolor}
\newcommand{\method}{RAMP}
\newcommand{\methodfull}{Regime-Aware Multiobjective Pre-optimization}

\begin{document}

\title{GPU-Accelerated Robust MA--RIS Pre-Optimization for Communication and Target Illumination}

\author{Tran Le Dung and Ho Viet Duc Luong%
\thanks{Tran Le Dung and Ho Viet Duc Luong are with the School of Information and Communications Technology (SoICT), Hanoi University of Science and Technology (HUST), Hanoi 100000, Vietnam (e-mail: dung.tl224964@sis.hust.edu.vn; luonghvd@soict.hust.edu.vn). (\textit{Corresponding author: Ho Viet Duc Luong}.) This research is funded by Hanoi University of Science and Technology (HUST) under project number T2025-TN-006.}%
}

\markboth{IEEE WIRELESS COMMUNICATIONS LETTERS}%
{Dung \MakeLowercase{\textit{et al.}}: Robust MA--RIS Pre-Optimization}

\maketitle

\begin{abstract}
Movable antennas (MAs) and reconfigurable intelligent surfaces (RISs) offer strong spatial control for integrated sensing and communication (ISAC), but repeatedly adapting MA positions and RIS phases to each channel realization can be impractical. This letter develops a GPU-accelerated robust pre-optimization framework that jointly selects slow-timescale MA geometry and RIS phases and returns a reusable Pareto set balancing expected worst-user rate and transmit-side worst-target illumination. The illumination objective measures directional transmitted power rather than complete echo-detection or parameter-estimation performance. The proposed \methodfull{} (\method) preserves regime-specific Pareto populations, regulates cross-regime mating using Pareto-survival feedback, and constructs a worst-regime robust archive. Numerical results demonstrate competitive robust Pareto quality against the considered evolutionary baselines while avoiding online physical re-optimization.
\end{abstract}


\section{Introduction}
\label{sec:introduction}

\IEEEPARstart{I}{ntegrated} sensing and communication (ISAC) is regarded as a key paradigm for future wireless networks, where radio-frequency hardware, spectrum, and transmitted waveforms are jointly exploited for data delivery and environmental sensing \cite{zhang2026integrated}. MAs and RISs provide additional spatial and electromagnetic degrees of freedom by adjusting the antenna geometry and programmable reflection responses, respectively \cite{zhu2024movable,lyu2025movable,van2024active}. Their joint deployment is therefore attractive for communication and sensing, but optimizing MA positions, RIS phases, transmit beamforming, and sensing-oriented radiation patterns leads to a large-scale, non-convex, and strongly coupled problem. Existing MA-enabled communication and ISAC designs commonly adapt antenna positions to instantaneous conditions \cite{lyu2025movable,wu2025movable}, while RIS-aided ISAC methods often rely on alternating, successive-convex, or manifold-based optimization \cite{van2024active,geng2025joint}. Such approaches can provide strong realization-specific performance, but repeated optimization may be impractical when channel acquisition, numerical optimization, RIS control, and mechanical movement must occur within a short operational interval. {Fluid antenna systems (FASs) provide another important form of spatial reconfigurability. Recent FAS studies have demonstrated gains in spatial selectivity, outage performance, and user fairness \cite{liu2026fasrsma,liu2026outage}. While FAS and MA share the broader goal of exploiting spatial degrees of freedom, the present work considers physically reconfigurable MA coordinates that are jointly pre-optimized with RIS phases and then fixed over a slow-timescale operational interval.}

An alternative is slow-timescale pre-optimization, where the physical configuration is optimized over representative propagation conditions and then reused online. Carefully pre-optimized irregular arrays can retain a substantial portion of the benefit of real-time MA adaptation while avoiding frequent movement \cite{irshad2025pre}, and GPU-accelerated evolutionary pre-optimization has recently been studied for MA--RIS communication systems under multiple propagation regimes \cite{luong2026evolutionary}. However, existing pre-optimization studies remain primarily communication-oriented and do not construct a reusable Pareto set exposing the tradeoff between communication quality and directional target illumination. In this work, the MA hardware remains physically reconfigurable, but its geometry and the RIS phases are selected at a slow timescale and fixed throughout an operational interval; thus, the online architecture behaves as a pre-optimized irregular array rather than a system performing per-coherence-time movement.

\begin{figure}[t]
\centering
\includegraphics[width=.9\columnwidth]{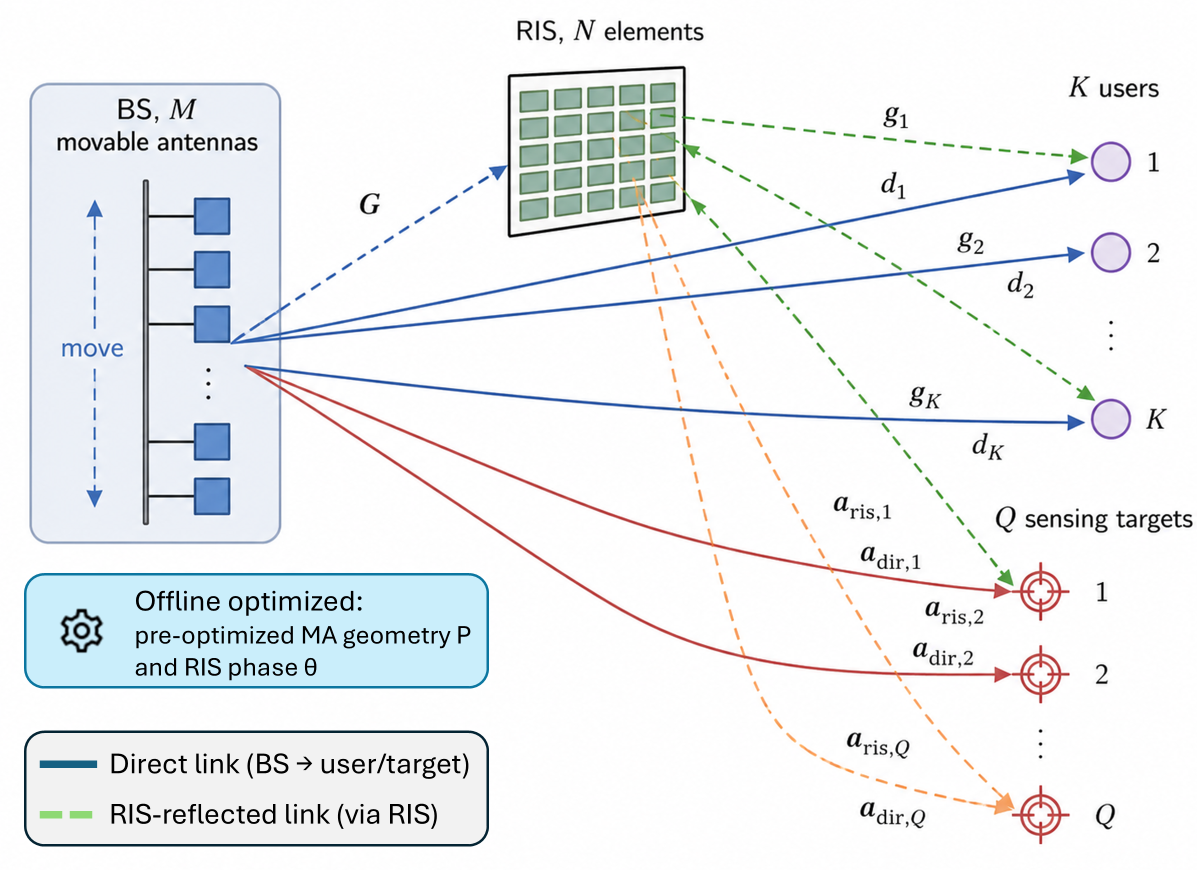}
\caption{MA--RIS downlink considered in this letter. The MA geometry $\mathbf P$ and RIS phase vector $\boldsymbol{\theta}$ are optimized offline and fixed during operation; only communication beamformers are recomputed online.}
\label{fig:system}
\end{figure}

This paper develops a GPU-accelerated robust multiobjective pre-optimization framework for joint MA geometry and RIS-phase design. A reusable set of MA--RIS configurations is optimized over stochastic user, channel, and target-direction realizations, balancing expected worst-user rate and expected worst-target transmit illumination. The illumination metric measures directional transmitted power rather than complete radar sensing performance; it excludes the return path, target radar cross section, clutter, receive processing, detection, and estimation. The proposed \method~preserves regime-specific Pareto populations, regulates cross-regime mating using Pareto-survival feedback, and re-evaluates archived candidates across all regimes to construct a reusable worst-regime robust Pareto set. 
The contributions are threefold. \textit{First}, we formulate a stochastic robust bi-objective pre-optimization problem for slow-timescale MA geometry and RIS-phase selection, balancing communication fairness and target illumination. \textit{Second}, we develop \method{}, combining regime-specific Pareto preservation, Pareto-survival-based transfer regulation, and worst-regime archive aggregation. \textit{Third}, we provide GPU-tensorized evaluation with statistical comparisons, transfer and regime-similarity analyses, runtime measurements, and sensitivity and scalability studies.
\section{System Model and Problem Formulation}
\label{sec:system_model}

A base station (BS) with $M$ MAs serves $K$ single-antenna users and illuminates $Q$ far-field target directions, assisted by an RIS with $L$ passive elements. We reserve $N_{\rm p}$ for the evolutionary population size. One environment realization $\Omega$ contains user positions, propagation regime, multipath coefficients, and target directions.

\subsection{Slow-Timescale Geometry and Communication Model}

The $m$-th MA is located at $\mathbf p_m=[x_m,y_m,z_m]^T$, and $\mathbf P=[\mathbf p_1,\ldots,\mathbf p_M]\in\mathbb R^{3\times M}$. The BS panel lies in the $y$--$z$ plane, so $x_m=x_0$ and only $(y_m,z_m)$ are controllable. Each MA moves inside a cell $\mathcal C_m$ and satisfies
\begin{equation}
 \mathbf p_m\in\mathcal C_m,\qquad
 \|\mathbf p_m-\mathbf p_{m'}\|_2\ge d_{\min},\quad m\ne m'.
 \label{eq:spacing_constraint}
\end{equation}
The RIS applies $\boldsymbol\Phi=\operatorname{diag}(e^{j\theta_1},\ldots,e^{j\theta_L})$, where $\theta_l\in[0,2\pi)$. The slow-timescale configuration is
\begin{equation}
 \mathbf x\triangleq(\mathbf P,\boldsymbol\theta).
 \label{eq:decision_vector}
\end{equation}

For user $k$, the effective channel row is
\begin{equation}
 \mathbf h_k^H(\mathbf x;\Omega)=\mathbf d_k^H(\mathbf P;\Omega)
 +\mathbf g_k^H(\Omega)\boldsymbol\Phi\mathbf G(\mathbf P;\Omega),
 \label{eq:effective_channel}
\end{equation}
where $\mathbf d_k^H\in\mathbb C^{1\times M}$, $\mathbf G\in\mathbb C^{L\times M}$, and $\mathbf g_k^H\in\mathbb C^{1\times L}$ are the BS--user, BS--RIS, and RIS--user channels. The implementation uses a position-aware field-response model for both line-of-sight (LoS) and multipath components.

Let $\mathbf H=[\mathbf h_1,\ldots,\mathbf h_K]^H$. For user $k$, define $\mathbf H_{-k}$ by deleting the $k$th row of $\mathbf H$. Assuming full row rank, let $\mathbf V_{0,k}$ contain the last $M-(K-1)$ right-singular vectors of $\mathbf H_{-k}$. The block-diagonalization beamformer is
\begin{equation}
 \widetilde{\mathbf w}_k=\mathbf V_{0,k}\mathbf V_{0,k}^H\mathbf h_k,
 \qquad
 \mathbf w_k=\frac{\widetilde{\mathbf w}_k}
 {\max(\|\widetilde{\mathbf w}_k\|_2,\epsilon)}.
 \label{eq:bd}
\end{equation}
For single-antenna users, this is the zero-forcing form of block diagonalization. The numerical floor is $\epsilon=10^{-12}$.
We use equal power $\rho_k=P_{\max}/K$ as a fixed online policy. With transmit error-vector-magnitude coefficient $\kappa$, the implemented signal-to-interference-plus-noise ratio is
\begin{equation}
 \gamma_k=\frac{(1-\kappa^2)\rho_k|\mathbf h_k^H\mathbf w_k|^2}
 {\sigma^2+\kappa^2\sum_{m=1}^{M}|h_{k,m}|^2
 \sum_{i=1}^{K}\rho_i|w_{m,i}|^2},
 \label{eq:sinr}
\end{equation}
and $R_k=\log_2(1+\gamma_k)$. Inter-user interference is nulled by \eqref{eq:bd} up to numerical precision. 

\subsection{Aperture-Normalized Target-Illumination Model}

For target direction $\varphi_q$, the direct BS steering vector is
\begin{equation}
 \mathbf a_{\rm d}(\mathbf P,\varphi_q)=
 \left[e^{j\frac{2\pi}{\lambda}\mathbf u^T(\varphi_q)\mathbf p_1},\ldots,
 e^{j\frac{2\pi}{\lambda}\mathbf u^T(\varphi_q)\mathbf p_M}\right]^T.
 \label{eq:direct_steering}
\end{equation}
Let $\mathbf a_{\rm r}(\varphi_q)\in\mathbb C^{L}$ be the RIS steering vector and $\mathbf G_{\rm geo}(\mathbf P)\in\mathbb C^{L\times M}$ the deterministic geometric BS--RIS response. We explicitly define the normalized quantities
 $\overline{\mathbf a}_{\rm r}={\mathbf a_{\rm r}}/{\sqrt L},
 \overline{\mathbf G}_{\rm geo}={\mathbf G_{\rm geo}}/{\sqrt L}$.
The effective transmit response is
\begin{equation}
 \mathbf a_{\rm eff}(\mathbf x;\varphi_q)=\mathbf a_{\rm d}(\mathbf P,\varphi_q)
 +\eta_q\overline{\mathbf G}_{\rm geo}^{T}(\mathbf P)
 \boldsymbol\Phi\overline{\mathbf a}_{\rm r}(\varphi_q).
 \label{eq:a_eff}
\end{equation}
Under this chosen unit-norm convention, the reflected field contains the corresponding effective $1/L$ factor. This is an aperture-normalized modeling convention used to isolate phase-control effects; it is not asserted as a physical scaling law for received echo power as $L$ changes. The dimensionless coefficient $\eta_q$ represents the relative reflected-path field strength after the chosen reference-link normalization, including BS--RIS and RIS--target attenuation and reflection efficiency. The default experiment uses $\eta_q=\eta=1$ for all target directions. RIS elements are otherwise modeled as ideal unit-modulus reflectors; non-ideal amplitude loss is not modeled separately and may be absorbed into $\eta_q$. When $L$ is varied, Sec.~\ref{sec:sensitivity} keeps half-wavelength inter-element spacing, so the physical aperture grows with $L$ while the same normalization convention is retained.

With $\mathbf R_s=\sum_{k=1}^{K}\rho_k\mathbf w_k\mathbf w_k^H$, the transmit illumination toward target $q$ is
\begin{equation}
 B_q(\mathbf x;\Omega)=\mathbf a_{\rm eff}^H\mathbf R_s\mathbf a_{\rm eff},
 \quad B_{\min}(\mathbf x;\Omega)=\min_q B_q(\mathbf x;\Omega).
 \label{eq:illumination}
\end{equation}
Equation~\eqref{eq:illumination} measures directional transmitted power and excludes two-way propagation, radar cross section, clutter/noise, receive processing, detection, and estimation; target range is therefore not modeled.

\subsection{Stochastic Robust Multiobjective Design}

The two objectives are
\begin{align}
 F_{\rm com}(\mathbf x)=\mathbb E_{\Omega}\!\left[\min_k R_k(\mathbf x;\Omega)\right],
 F_{\rm ill}(\mathbf x)=\mathbb E_{\Omega}\!\left[B_{\min}(\mathbf x;\Omega)\right].
 \label{eq:objectives}
\end{align}
For a Monte Carlo batch $\{\Omega_b\}_{b=1}^{B}$, both expectations are replaced by sample means. The offline problem is
\begin{subequations}
\label{eq:main_problem}
\begin{align}
 \max_{\mathbf x=(\mathbf P,\boldsymbol\theta)}\quad
 &\mathbf F(\mathbf x)=[F_{\rm com}(\mathbf x),F_{\rm ill}(\mathbf x)]\\
 \text{s.t.}\quad &\eqref{eq:spacing_constraint},\qquad \theta_l\in[0,2\pi),\;l=1,\ldots,L.
\end{align}
\end{subequations}
For propagation regime $t$, denote the regime-specific objectives by $\mathbf F^{(t)}$. The deployable robust values are defined as
\begin{equation}
 F_j^{\rm rob}(\mathbf x)=\min_{t=1,\ldots,T}F_j^{(t)}(\mathbf x),
 \qquad j\in\{\mathrm{com},\mathrm{ill}\}.
 \label{eq:robust}
\end{equation}

\section{Regime-Aware Multiobjective Pre-Optimization}
\label{sec:proposed_method}

\subsection{Representation, Selection, and Adaptive Transfer}

The normalized genotype is $\mathbf z\in[0,1]^D$, where $D=2M+L$. The first $2M$ variables encode in-plane MA offsets and the remaining $L$ variables encode RIS phases. Decoding uses bounded soft clipping and a repulsive repair for \eqref{eq:spacing_constraint}.
Each individual has a skill factor $t$. Environmental selection is carried out separately within each regime using standard fast non-dominated sorting and crowding distance \cite{deb2002fast}. Cross-regime mating is attempted with random-mating probability $r_g$. The proposed adaptation signal is based on whether transfer offspring survive into their assigned regime's rank-one front. Let $s_g^{\rm tr}$ and $s_g^{\rm in}$ be the generation-$g$ survival rates of transfer and intra-regime offspring. Their exponential moving averages are
\begin{equation}
 \bar s_g^a=(1-\beta)\bar s_{g-1}^a+\beta s_g^a,
 \qquad a\in\{\mathrm{tr},\mathrm{in}\},
 \label{eq:survival_ema}
\end{equation}
\begin{equation}
 r_{g+1}=\operatorname{clip}\!\left(r_g+\alpha(\bar s_g^{\rm tr}-\bar s_g^{\rm in}),
 r_{\min},r_{\max}\right).
 \label{eq:rmp_update}
\end{equation}
Hence, cross-regime mating increases only when its Pareto-survival rate exceeds the contemporaneous intra-regime baseline. This Pareto-conditioned transfer controller, not the inherited
crossover or NSGA-II operators, is the algorithmic distinction
to the communication-oriented framework
in \cite{luong2026evolutionary}.

\subsection{Archive and Deployable Robust Front}

Unlike pooled NSGA-II, RAMP preserves per-regime fronts; unlike independent NSGA-II, it also uses survival-driven transfer. Every regime-specific rank-one configuration is stored in an external archive $\mathcal A$. At termination, all archived configurations are re-evaluated on every regime using a common evaluation set. The algorithm applies \eqref{eq:robust} and returns the non-dominated set $\mathcal A_{\rm rob}$.
{At a deployment, calibration, or other slow-timescale reconfiguration boundary, one operating point is selected from $\mathcal A_{\rm rob}$ according to the desired communication--illumination tradeoff. Its associated MA geometry and RIS phase vector are then fixed throughout the subsequent operational interval. For each channel realization within that interval, only the communication beamformers are recomputed; the system does not switch among Pareto configurations on a per-realization basis. Switching to another Pareto point requires a new slow-timescale physical reconfiguration.}

\begin{algorithm}[t]
\caption{RAMP robust pre-optimization}
\label{alg:ramp}
\scriptsize
\begin{algorithmic}[1]
\REQUIRE $N_{\rm p},G,B$, regimes $1{:}T$, $r_0,r_{\min},r_{\max},\alpha,\beta$
\ENSURE Unified robust Pareto set $\mathcal A_{\rm rob}$
\STATE Initialize genotypes, balanced skill factors, and $\mathcal A\leftarrow\emptyset$.
\FOR{$g=1$ to $G$}
 \STATE Draw a shared Monte Carlo batch for generation $g$.
 \STATE Generate intra-/cross-regime offspring using $r_g$.
 \STATE Decode, repair, and GPU-evaluate on the assigned regime.
 \STATE Select per regime; update $r_g$ from transfer/intra rank-one survival.
 \STATE Add regime-specific rank-one genotypes to $\mathcal A$.
\ENDFOR
\STATE Re-evaluate $\mathcal A$ on all regimes and apply \eqref{eq:robust}.
\RETURN non-dominated robust set $\mathcal A_{\rm rob}$.
\end{algorithmic}
\end{algorithm}

For clarity, Table~\ref{tab:positioning} summarizes the positioning of RAMP against the communication-oriented pre-optimization framework, referred to as MGRAP, in \cite{luong2026evolutionary}, pooled NSGA-II, and independent NSGA-II in terms of regime handling, transfer signal, and robust-set
construction.

\begin{table}[t]
\centering
\caption{Positioning against representative pre-optimization and search strategies.}\label{tab:positioning}
\scriptsize
\setlength{\tabcolsep}{3pt}
\begin{tabular}{lccc}
\toprule
Approach & Per-regime & Transfer signal & Robust set\\
\midrule
MGRAP pre-opt. \cite{luong2026evolutionary} & Yes & Scalar & No\\
Pooled NSGA-II & No & None & No\\
Independent NSGA-II & Yes & None & Yes\\
RAMP & Yes & Pareto survival & Yes\\
\bottomrule
\end{tabular}
\end{table}

\subsection{GPU Evaluation and Complexity}

For a candidate and one regime, the principal costs over $B$ Monte Carlo samples are channel construction, batched singular value decompositions for \eqref{eq:bd}, rate evaluation, and target illumination. Let $P_{\rm sc}$ be the number of multipath components. A conservative per-candidate cost is
\begin{equation}
 \mathcal O\!\left(B\left[KMP_{\rm sc}+LM+KLM+MK^3+QMK\right]\right).
 \label{eq:complexity_eval}
\end{equation}
The $KLM$ term is the cascaded RIS communication channel, $MK^3$ represents $K$ SVD/null-space constructions for $(K-1)\times M$ matrices, and $QMK$ is the illumination projection. For balanced regime populations of sizes $n_t$, the evolutionary stage costs
\begin{equation}
 \mathcal O\!\left(GN_{\rm p} C_{\rm eval}+G\sum_{t=1}^{T}n_t^2\right),
 \label{eq:complexity_total}
\end{equation}
for the optimization itself, while final archive evaluation adds $\mathcal O(|\mathcal A|TC_{\rm eval})$. The best-so-far convergence tracker used only for reporting additionally re-evaluates the current rank-one union on all regimes. GPU batching does not change these asymptotic orders; it parallelizes the candidate and Monte Carlo axes and reduces wall-clock time. The implementation chunks the archive evaluation to control peak memory.

\section{Numerical Results}
\label{sec:numerical_results}

\subsection{Settings, Reproducibility, and Statistical Protocol}

Table~\ref{tab:parameters} gives the settings. Table~\ref{tab:quality} uses $15$ paired seeds, common environments and a common final-evaluation set, with HV reference $(0,-30\,\mathrm{dB})$. Illumination is evaluated in linear scale inside the optimizer and expressed in dB for the reported fronts and Pareto-quality metrics. For IGD, the non-dominated solutions from the $15$ seeds are merged for each method, and the common reference front is the non-dominated union of the five merged method fronts. Paired $95\%$ confidence intervals (CIs) are formed from the per-seed differences $\Delta\mathrm{HV}$ and reported in Table~\ref{tab:quality}.

\begin{table}[t]
\centering
\caption{Main simulation and optimization settings.}
\label{tab:parameters}
\scriptsize
\setlength{\tabcolsep}{3pt}
\begin{tabular}{p{0.36\columnwidth}p{0.56\columnwidth}}
\toprule
Parameter & Value\\
\midrule
Carrier; $M,K,Q$; RIS & $28$ GHz; $16,4,3$; $8\times8$ ($L=64$)\\
MA/RIS geometry & $4\times4$ cells; cell pitch $\lambda$; cell extent $\lambda/2$; $d_{\min}=\lambda/2$; RIS $(30,40,10)$m, spacing $\lambda/2$\\
Users / target directions & users at $50$--$200$ m in a $60^\circ$ sector; far-field target directions with $20^\circ/20^\circ$ offset/spread\\
Channel & $61.4+22\log_{10}d$ dB; no shadowing; $4$ paths, $120^\circ$ spread; Rician $6$ dB\\
Power/noise/EVM/RIS & $30$ dBm; $180$ kHz; $9$ dB; $\kappa=0.05$; unit-modulus RIS; $\eta=1$\\
Optimizer & $N_{\rm p}=200$, $G=300$, $B=32$; $r_0=0.3$, $[0.05,0.5]$, $\alpha=0.05$, $\beta=0.1$\\
Seeds & $15$ paired seeds for each reported comparison\\
\bottomrule
\end{tabular}
\end{table}

Experiments used an RTX~5070~Ti (16~GB; driver 591.86), Python~3.14.3, PyTorch~2.11.0+cu128/CUDA~12.8, and an i9-14900K with 64~GB RAM. 
The implementation is publicly available at {\url{https://github.com/gittbin/RAMP}}. {Performance is evaluated on the unified robust, worst-regime Pareto front. Hypervolume (HV) measures the objective-space region dominated by an approximation front relative to the fixed reference point; larger values indicate better convergence and coverage. Inverted generational distance (IGD) is the average distance from the common reference front to its nearest points in the set; smaller values are better. The evolutionary baselines include multi-objective differential evolution (MO-DE), which uses differential operators with Pareto-based selection, and multi-objective particle swarm optimization (MOPSO), which uses swarm updates guided by particle and archive leaders.}

\subsection{Pareto Quality and Transfer Effectiveness}

\begin{figure}[t]
\centering
\includegraphics[width=\columnwidth]{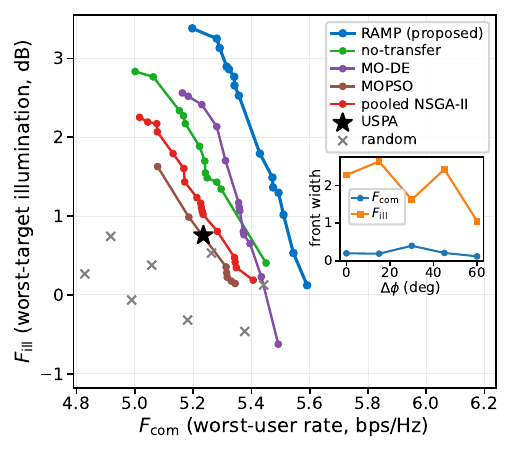}
\caption{Unified worst-regime fronts; the second axis is transmit target illumination. HV reference: $(0,-30\,\mathrm{dB})$. $\text{HV}_{\text{RAMP}}=186.05$.}
\label{fig:front}
\end{figure}

{Figure~\ref{fig:front} uses the paired seed whose RAMP HV is closest to the $15$-seed mean. The displayed RAMP front spans approximately $F_{\rm com}\in[5.197, 5.591]$~bps/Hz and $F_{\rm ill}\in[0.121 , 3.381]$ dB, providing a broad communication--illumination tradeoff and higher illumination than the fixed-array point and most sampled random layouts at comparable communication rates.}

\begin{table}[t]
\centering
\caption{Robust-front quality over $15$ paired seeds. CI is for $\Delta$HV = RAMP $-$ method.}
\label{tab:quality}
\scriptsize
\setlength{\tabcolsep}{4.2pt}
\begin{tabular}{lccc}
\toprule
Method & HV & IGD & $95\%$ CI\\
\midrule
\method{} (adaptive) & $\mathbf{182.59\pm19.1}$ & $\mathbf{0.018}$ & --\\
No transfer ($r=0$) & $178.23\pm20.0$ & $0.066$ & $[3.17,5.55]$\\
MO-DE & $178.16\pm19.9$ & $0.128$ & $[2.79,6.07]$\\
Fixed transfer ($r=0.3$) & $176.45\pm19.8$ & $0.150$ & $[4.71,7.63]$\\
Pooled & $174.86\pm20.3$ & $0.758$ & $[5.56,9.90]$\\
MOPSO & $167.71\pm17.2$ & $1.434$ & $[12.34,17.42]$\\
\bottomrule
\end{tabular}
\end{table}

Table~\ref{tab:quality} ranks RAMP first in mean HV and IGD. Its mean HV gains are $4.36$, $4.43$, $6.14$, $7.73$, and $14.88$ over no transfer, MO-DE, {fixed transfer}, pooled, and MOPSO, respectively. All reported paired $95\%$ CIs remain above zero.

\begin{figure}[t]
\centering
\includegraphics[width=.9\columnwidth]{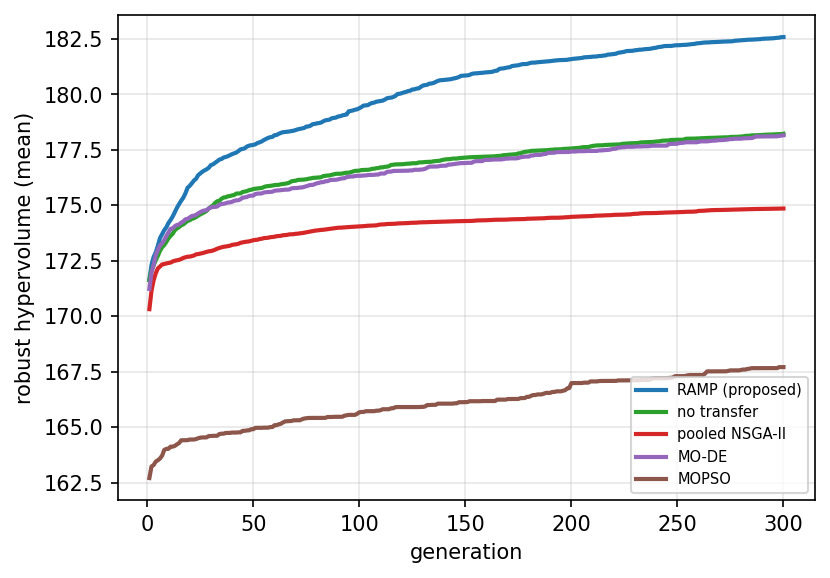}
\caption{Mean best-so-far robust HV over the $15$ paired seeds.}
\label{fig:hv}
\end{figure}

Across $500$ shared configurations, median regime-pair Spearman
correlations are $0.59$ for $F_{\rm com}$ and $0.53$ for
$F_{\rm ill}$. {Under the same $15$ paired seeds,
optimization budget, archive construction, and common final-evaluation
set, adaptive RAMP achieves a mean HV of $182.59$, compared with
$176.45$ for fixed transfer ($r=0.3$) and $178.23$ with no
transfer ($r=0$). The lower HV of fixed transfer relative to
no transfer is consistent with negative transfer: maintaining a
constant cross-regime mating rate can introduce offspring that are
less useful to the receiving regime when inter-regime similarity is
only moderate. In contrast, adaptive RAMP outperforms both
variants, indicating that cross-regime knowledge can be beneficial
when transfer is selectively regulated according to Pareto-survival
feedback.} For adaptive RAMP, the controller ends at
$r_G=0.060$; transfer rank-one survival ($0.18$) is below
intra-regime survival ($0.23$), {consistent with
reducing cross-regime mating when transfer becomes less productive.}
\subsection{Runtime, Sensitivity, and Scalability}
\label{sec:sensitivity}

\begin{table}[t]
\centering
\caption{Measured computational cost at the default setting.}
\label{tab:runtime}
\scriptsize
\setlength{\tabcolsep}{2.5pt}
\begin{tabular}{lrrr}
\toprule
Method & Time (s) & Memory (MB) & Eval./s\\
\midrule
RAMP & $80.0$ & $3822$ & $1845$\\
No transfer & $76.6$ & $3822$ & $1941$\\
MO-DE & $48.8$ & $318$ & $2805$\\
Pooled & $56.3$ & $2239$ & $2726$\\
MOPSO & $45.8$ & $366$ & $1587$\\
\midrule
\multicolumn{4}{@{}l@{}}{Matched GPU speedup: $(4.12\pm0.20)\times$ total; $3.09\times$ search-only.}\\
\bottomrule
\end{tabular}
\end{table}

At the default setting, GPU tensorization accelerates RAMP by $(4.12\pm0.20)\times$ in total wall-clock time and by $3.09\times$ for the evolutionary search stage. RAMP requires only $4.4\%$ more total time than no transfer; its $3.8$-GB peak allocation occurs during the one-off archive re-evaluation, which accounts for $15.4\%$ of total time and remains well within the 16-GB GPU capacity. Because RAMP is executed offline, this overhead does not affect online latency. Table~\ref{tab:runtime} summarizes the measured computational cost of the considered methods under their default experimental settings.

Sensitivity is evaluated one factor at a time in separate auxiliary $15$-seed sweeps; sweep-specific IGD references are used, so IGD values are compared only within the same sweep. The auxiliary HV values are the final values of the fixed-set convergence tracker used inside each sweep, whereas Table~\ref{tab:quality} reports archived robust fronts after common post-hoc final re-evaluation. Consequently, the auxiliary default value $190.8$ and the main-table mean $182.59$ are not estimates from the same evaluation set and should not be compared in absolute magnitude. For $B\in\{8,16,32,64\}$, the mean tracked HV values are $193.0$, $192.5$, $190.8$, and $191.5$, with standard deviations decreasing from $37.6$ to $14.3$ as Monte Carlo noise is reduced; runtime increases to $115$~s at $B=64$. For $L\in\{32,64,128\}$ under fixed half-wavelength spacing and the adopted unit-norm RIS convention, mean tracked HV is $195.5$, $190.8$, and $186.3$. This fixed-budget trend reflects both the increased decision dimension and the chosen normalization and should not be interpreted as a general physical disadvantage of a larger RIS. For $M\in\{8,16,24\}$, mean tracked HV is $38.7$, $190.8$, and $206.5$, indicating the importance of available spatial degrees of freedom under the fixed search budget. Peak memory across these sweeps ranges from $1424$ to $3827$~MB. Under the additional inverse-effective-gain policy
\begin{equation}
    \rho_k\propto\frac{1}{\max(|\mathbf h_k^H\mathbf w_k|^2,\epsilon_{\rm p})},
 \qquad \sum_k\rho_k=P_{\max},
\end{equation}
with $\epsilon_{\rm p}=10^{-12}$, the auxiliary tracked mean HV for RAMP changes from $190.8\pm20.5$ to $166.5\pm17.7$, while the ordering is preserved under both policies.

\subsection{Offline Reuse Versus Per-Realization Optimization}

\begin{figure}[t]
\centering
\includegraphics[width=\columnwidth]{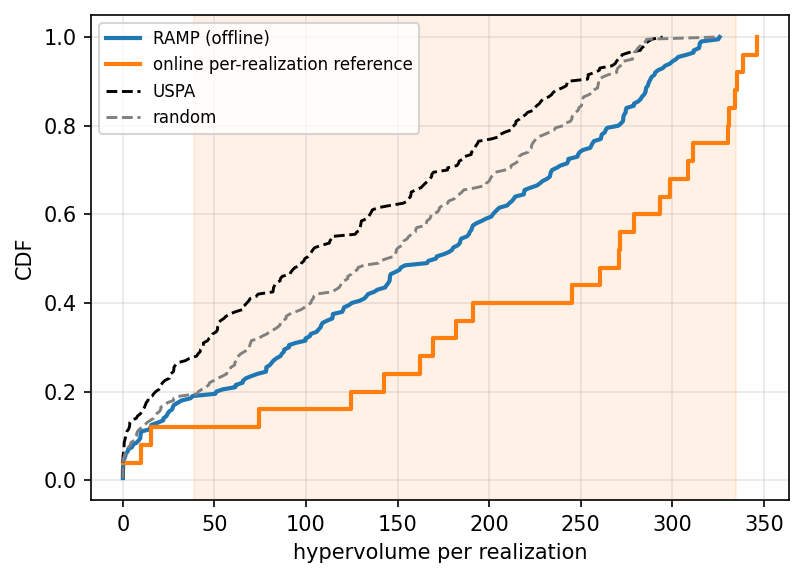}
\caption{Per-realization HV CDF using $200$ offline and $30$ online-reference realizations, with $75$ generations per online run. The shaded region denotes the $10$--$90\%$ interval of the online reference.}
\label{fig:cdf}
\end{figure}

Figure~\ref{fig:cdf} compares the fixed offline set with a per-realization optimization reference. The latter is expected to perform better because it uses realization-specific optimization, but it is not a mathematical upper bound because it is itself a finite-budget heuristic. The median per-realization HV is $175.3$ for the offline design and $265.3$ for the online reference. Selecting an operating point from the stored offline front takes $47.7\pm10.0$~ms, whereas a full per-realization re-optimization takes $11.14\pm1.71$~s. These timings correspond to different online operations, but they quantify the latency saved by reusing the precomputed front rather than solving the high-dimensional layout problem again.

\section{Conclusions}
\label{sec:conclusion}

This letter presented \method{}, a GPU-tensorized robust pre-optimization framework for slow-timescale MA geometry and RIS-phase design. It constructs a reusable worst-regime Pareto set balancing expected worst-user rate and transmit-side worst-target illumination, with the latter treated explicitly as an illumination proxy rather than a complete radar metric. RAMP combines regime-specific Pareto preservation, Pareto-survival-based transfer regulation, and common cross-regime archive re-evaluation; the adaptive transfer mechanism suppresses unproductive cross-regime exchange and  {thereby mitigates negative transfer}. Across the evaluated baselines, RAMP achieves the highest mean robust HV and the lowest IGD, with paired confidence intervals supporting the HV improvements. The GPU implementation requires $80.0$~s at the default setting and provides a $(4.12\pm0.20)\times$ speedup over the matched CPU implementation. The current formulation is limited to far-field transmit illumination, a fixed online power policy, slow-timescale MA reconfiguration, and normalization-dependent RIS-size behavior. Future work will extend the framework to two-way sensing models, receive processing, estimation-oriented objectives, and joint power allocation.

\bibliographystyle{IEEEtran}
\bibliography{reference}

\end{document}